# Experiences in Implementing an ICT-Augmented Reality as an Immersive Learning System for a Philippine HEI

Nestor R. Valdez[1], Marcelo V. Rivera[1] and Jaderick P. Pabico[2]

[1]*Technological University of the Philippines–Taguig, Taguig City 1632, Metro Manila*
[2]*Institute of Computer Science, University of the Philippines Los Baños, College 4031, Laguna*

[1]{redhatpoinks,marcvrivera}@gmail.com, [2]jppabico@uplb.edu.ph

**Abstract** – *The use and implementation of virtual learning environments (VLEs) in higher education instruction has been in place since the advent of computer networking and popularized by the accessibility of the Internet. Because of the fast-paced advances in hardware and software technologies that stimulate the human senses, the number and the quality of researches on the use of three-dimensional (3D) environments to augment the experiences of learners have increased in recent years. Since VLEs can be customized for the learners' needs, they have been used to provide a virtual immersion that is otherwise difficult, if not impossible, to experience in real-world (e.g., teaching the effect of varying planetary gravities on objects).*

*This paper presents the experiences in building and implementing a 3D avatar-based virtual world (3D-AVW) as a VLE (3D-AVLE) for the Technological University of the Philippines-Taguig (TUP-T), a higher education institution (HEI) in the Philippines. Free and Open Source Software (FOSS) systems were used, such as the OpenSimulator and various 3D renderers, to create a replica of the TUP-T campus in a simulated 3D world. The 3D-AVLE runs in a single server that is connected to the learners' computers via a simply-wired local area network (LAN). The use of various networking optimization techniques was experimented on to provide the learners and the instructors alike a seamless experience and lag-less immersion within the 3D-AVLE. With the current LAN setup in TUP-T, the optimal number of concurrent users that can be accommodated without sacrificing connectivity and the quality of virtual experience was found to be at 30 users, exactly the mean class size in TUP-T. The 3D-AVLE allows for recording of the learners' experiences which provides the learners a facility to review the lessons at a later time. Classes in fundamental topics in engineering sciences were conducted using the usual teaching aid technologies such as the presentation software, and by dragging-and-dropping the presentation files to the 3D-AVLE, resulting to increased learning curve by both the instructors and the learners.*

## INTRODUCTION

The interest of students to latest information and communication technology (ICT) applications and tools, particularly online and network games, multimedia, social networking, chat, and Internet surfing, has been recently persevering. Students consuming excessive hours on different online games and activities have always been a concern to parents as well as educators because these seemingly "addictive" technological innovations are "taking away students from books towards electronic gadgets" [1]. There is an ongoing pattern that the classical way of teaching is no longer effective to today's technology-aware youth. To bring back the youth to "reading books," a management change that use these technological innovations for instruction is inevitable [2]. More and more educational institutions worldwide have already found out that resistance to the forces of change brought about by ICT is a futile curriculum management strategy. Educational institutions have no other recourse but to accept these changes and adapt to the innovations offered by these technologies [3].

The recent developments in ICT must not be seen by educators and school administrators as its competitors to students' attention but as opportunities to evaluate the prospect of these developments as areas to enhance the students' learning experience. Experiences learned by educators worldwide in using such technologies show that there are a plethora of motivating factors in using

ICT for enhanced learning. One among the many cited motivations is the flexibility of these technologies to adapt to the learners' needs. The potential of ICT to augment the traditional classroom settings has already been exploited by others in immersing students, for example, to the reconstruction of a combination of real and imaginary environments that are remotely located, existed in different times, and scaled differently [4]. Imagine the experience a learner would have in a world geography class if she*can be in a foreign land complete with the reconstruction of its geographical features, or in a history class in an ancient site complete with artifacts, or in a chemistry class in a sophisticated laboratory and is able to handle contraptions and chemicals safely, or in an astronomy class in a galaxy reduced to a navigable size. This list can go on and on without boundary because there is no limit to learning with ICT-augmented reality.

To date, there exists a number of developed systems that provide augmented reality using 3D avatar-base virtual worlds (3D-AVW). Examples of such systems are the (Teen) Second Life (SL/TSL) [5], Open Wonderland (OWL) [6], and Open Simulator (OpenSim) [7]. SL/TSL is available commercially while OWL and OpenSim are free and open source software (FOSS) systems. Some higher education institutions (HEIs) worldwide have already implemented these 3D-AVW as their own VLE resulting to a 3D avatar-based virtual learning environment (3D-AVLE). Figure 1 shows an example screenshot of one learning session using SL/TSL.

To date, there are no studies that have been conducted in the Philippines that measure the perception and performance of the students in a 3D-AVLE. This research endeavor aims to provide answers to the following questions: How this new teaching approach will affect the academic performance of the students? Will this method be more effective than the traditional classroom type of teaching? With dissimilarity in culture compared to other countries, as well as the various regional differences within the country, it is most likely that the result will vary if conducted in a national level. Although the end result of this research undertaking is to measure the performance of students who were taught certain topics under the 3D-AVLE, the various outcomes of the research are presented in three different parts: software technologies, hardware setup optimization, and classroom implementation experiences; student perception; and student performance.

This paper presents the first part in the hope that other educational institutions within the Philippine setting will learn from the experiences and come up with a better or customized 3D-AVLE setup of their own. The other two parts are discussed in detail in separate presentations [8,9].

With the often and vast changes in ICT, it is important to be in line with the current trend, particularly with the 3D-AVLE. As far as the records show in various archiving publications, this research study was the first to be conducted within the Technological University of the Philippines-Taguig (TUP-T), if not in the whole higher education system in the country. Here, the students' perception and performance of teaching and learning while being conducted in 3D-AVLE were measured. With this implementation, it is envisioned that the TUP-T campus will benefit in the future with the development of 3D-AVLE for TUP-T.

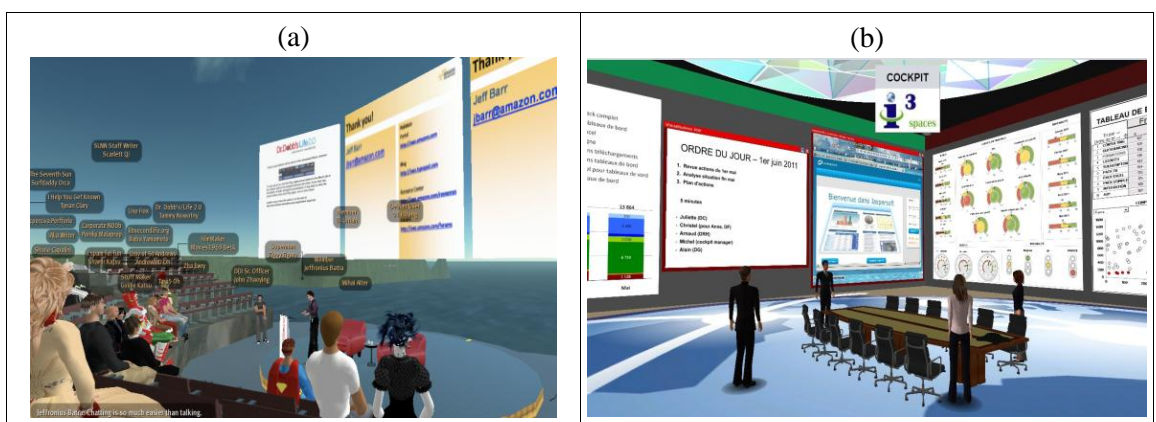

**Figure 1**. A screenshot of an example learning session in the point of view of one of the participants using (a) SL/TSL (image source: http://sitearm.files.wordpress.com), and (b) OWL (image source: http://www.hypergridbusiness.com).

Particularly, the curricula could be improved into using this new mode of teaching, and if successful may be able to accommodate the growing quantity of incoming students to TUP-T, not only to accommodate the growing population of the City of Taguig but also to the predicted influx of foreign students due to the formal implementation of the ASEAN Free Trade Area (AFTA) [10].

In this paper, the software technologies used in developing the 3D-AVLE are presented. The methodology used in optimizing the limited hardware systems that TUP-T allowed to be used for this purpose is shown. The experiences in implementing the 3D-AVLE as a different learning environment for the students are discussed. Just to provide completeness to the paper, a summarized portion of the results obtained in measuring the student perception and performance are also presented, although the details are presented elsewhere [8,9] and further improvements of these will be presented in the future.

**SOFTWARE TECHNOLOGIES**

We discuss in this section the various software technologies used in implementing the 3D-AVLE. Although we are able to find an acceptable performance of the 3D-AVLE with the combinations of these software technologies, we provide caution to the readers that these may not provide optimal student experience with their implementations as the system is also dependent on the variability brought about by the underlying hardware systems, data communication systems, and more importantly, the peopleware. Figure 2 shows the conceptual relationship of various software technologies used in 3D-AVLE. The 3D-AVLE is composed of a server which allows connections by the clients. The whole client-server connectivity is provided by OpenSim [7]. The virtual objects that are stored and computed by the server are rendered by the client and shown through a virtual environment viewer, while the audio communication between clients are handled by a standard audio transmission codec.

*A. 3D-AVLE*

One of the primary needs in creating a 3D-AVLE is the creation of 3D objects into the well-known SL/TSL. SL/TSL is the most popular commercially available 3D-AVLE platform that has been used by various open and distance education institutions worldwide. During the development days of SL/TSL, people would reproduce the buildings and classrooms they were familiar with in real life [11]. Within SL/TSL, university teachers typically think first of opportunities to present information in ways they know, such as with slides and video clips. While slides and video clips are possible in SL/TSL, it is more efficient to use other well-developed e-platforms intended for presentations. According to a SL/TSL researcher in an HEI setting, SL/TSL allowed them to explore potential applications for education, which were both impressive and surprising to everyone involved in the endeavour [12]. Students who participated in SL/TSL achieved a grade standing 28% higher than the previous class who did not utilize virtual worlds [13].

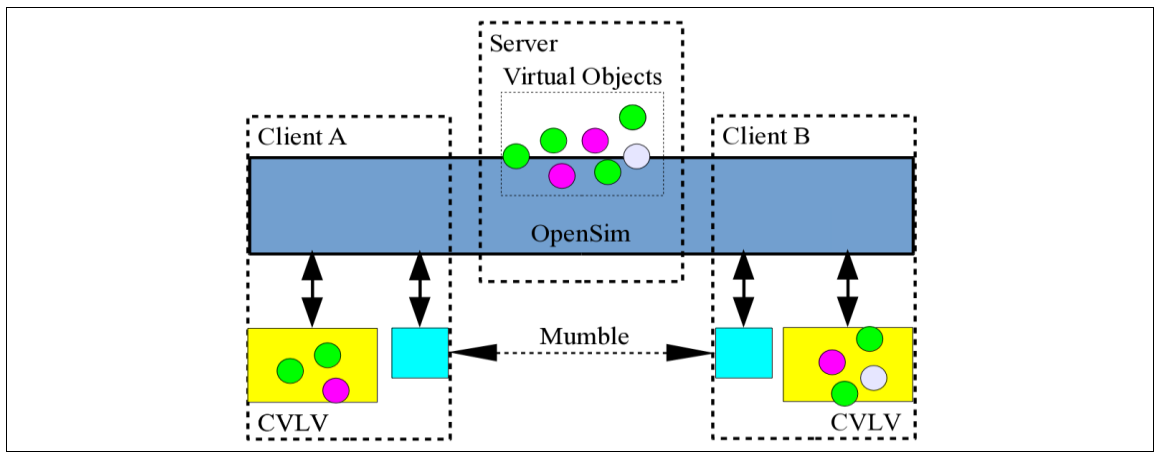

**Figure 2**. The conceptual relationship of various software technologies in 3D-AVLE.

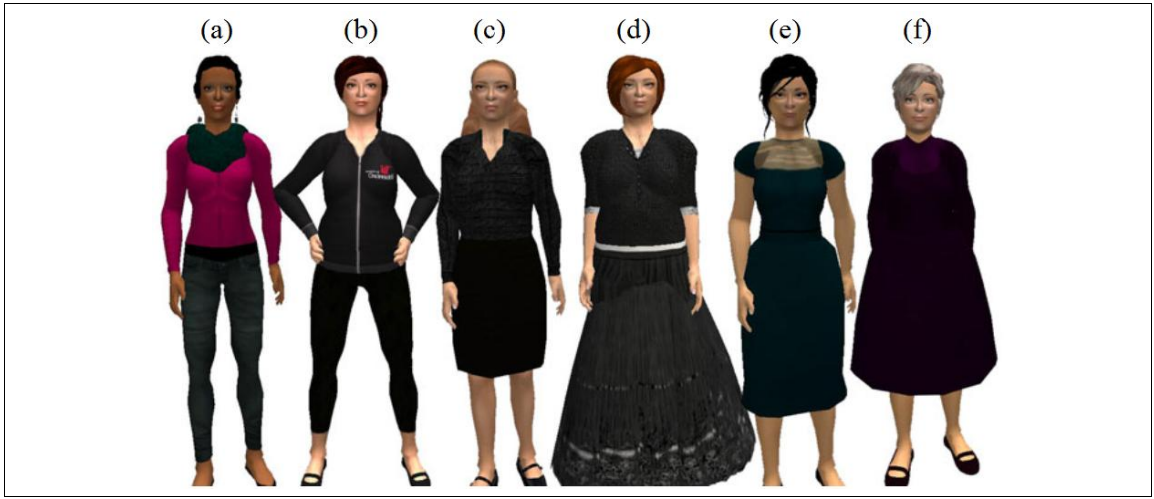

**Figure 3**. An age-group selection of avatars for the female gender in 3D-AVLE: (a) young adult, (b) collegiate, (c) graduate, (d) adult, (e) middle-aged, and (f) elderly.

In creating 3D-AVLEs, there are development platforms that can be used in order to lessen the cost, or even to freely make them. OpenSim [7] is a free and open source software (FOSS) 3D application server that can be run on multiple operating systems and hardware platforms. Because OpenSim uses the same messaging protocol which is the core of SL [5], it can be used to simulate virtual environments similar to SL/TSL [7]. However, since OpenSim is FOSS while SL/TSL is a commercial platform [14], then OpenSim is a more logical choice between the two for implementation in TUP-T.

*B. Avatars as Student Representatives*

3D-AVLE is a first-person point-of-view (FP-POV) system wherein what the student users see in the computer monitor is the immediate virtual environment. However, to differentiate it from most FP-POV systems, the 3D-AVLE has a capability of a perceived camera which pans-out of the point-of-view of the student user, giving the user a third-person point-of-view of her virtual representative. This virtual representative is called an avatar, a personalized graphic representation or rendering that represents a computer user or the user's alter ego or character. The avatar allows the 3D-AVLE user the capability to change roles: as an active actor as herself, and as a spectator. 3D-AVLE provides various capabilities to give the avatar its own personality. Figure 3 shows an example selection of avatars for the female gender in 3D-AVLE. There does exist a selection for the male gender also.

*C. Virtual Object Editors*

Another tool that we found essential in creating a virtual world is a virtual object editor, where another FOSS is the Sweet Home 3D (SH3D). SH3D is an interior design application software that helps in outlining the structure of classroom buildings and other 3D objects. SH3D is easier to use because it can import a blueprint of a floor plan which can speed up drawing of the walls of an existing structure [15]. However, SH3D saves work in a *wavefront* object file which is not known to OpenSim.

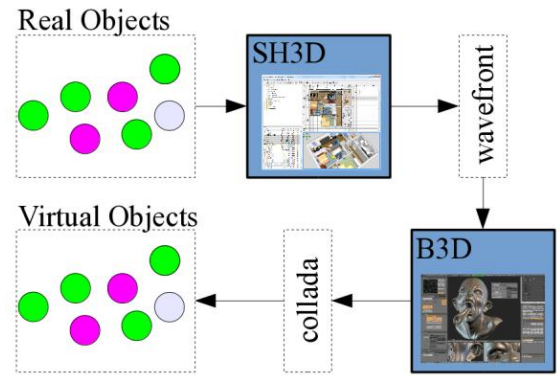

**Figure 4**. The flowchart for creating virtual objects using various virtual object editors.

Another editor that can be used in creating a 3D-AVLE is the FOSS Blender 3D (B3D) editor. B3D was developed to help graphics artists create 3D graphics [16-18]. In fact many professionals and 3D artists consider B3D as being the best open source solution for 3D computer graphics [19]. In this study, we used B3D to convert the *wavefront* object file into its *collada* counterpart. The *collada* formats are the 3D objects inputs for OpenSim. The flowchart with corresponding object editors used for creating virtual objects from models of real-world ones is shown in Figure 4.

*D. Virtual World Viewers*

In order for the 3D virtual objects to be visible and be manipulated by our students, we used the Cool Virtual Life Viewer (CVLV). CVLV is particularly used for SL/TSL and OpenSim grids [20], and is also FOSS. Another virtual life viewer is the Imprudence Viewer (IV), which is also being used worldwide with OpenSim. However, due to IV's limited functions like the capability to upload personalized objects [21], we chose to use the CVLV, instead.

With the CVLV, a student can manipulate the movement of her avatar using the combination of arrow key presses of the keyboard and the mouse movement, similar to how first-person shooter games are controlled. CVLV allows for photorealistic rendering of bumps in textures, of shadows due to the effect of trees and atmospheric phenomena, and of reflections on shiny objects. For example, bumps in virtual walls are rendered with photo-realism if the real-world walls are finished with bumps, multiple shadows are rendered correctly due to the effect of sunlight scattering as a result of cloudy skies, and wavy reflections are rendered correctly on water reflections. Figure 5 shows the TUP-T facade and the campus as respectively seen in photographs and as 3D-AVLE virtual objects through the CVLV.

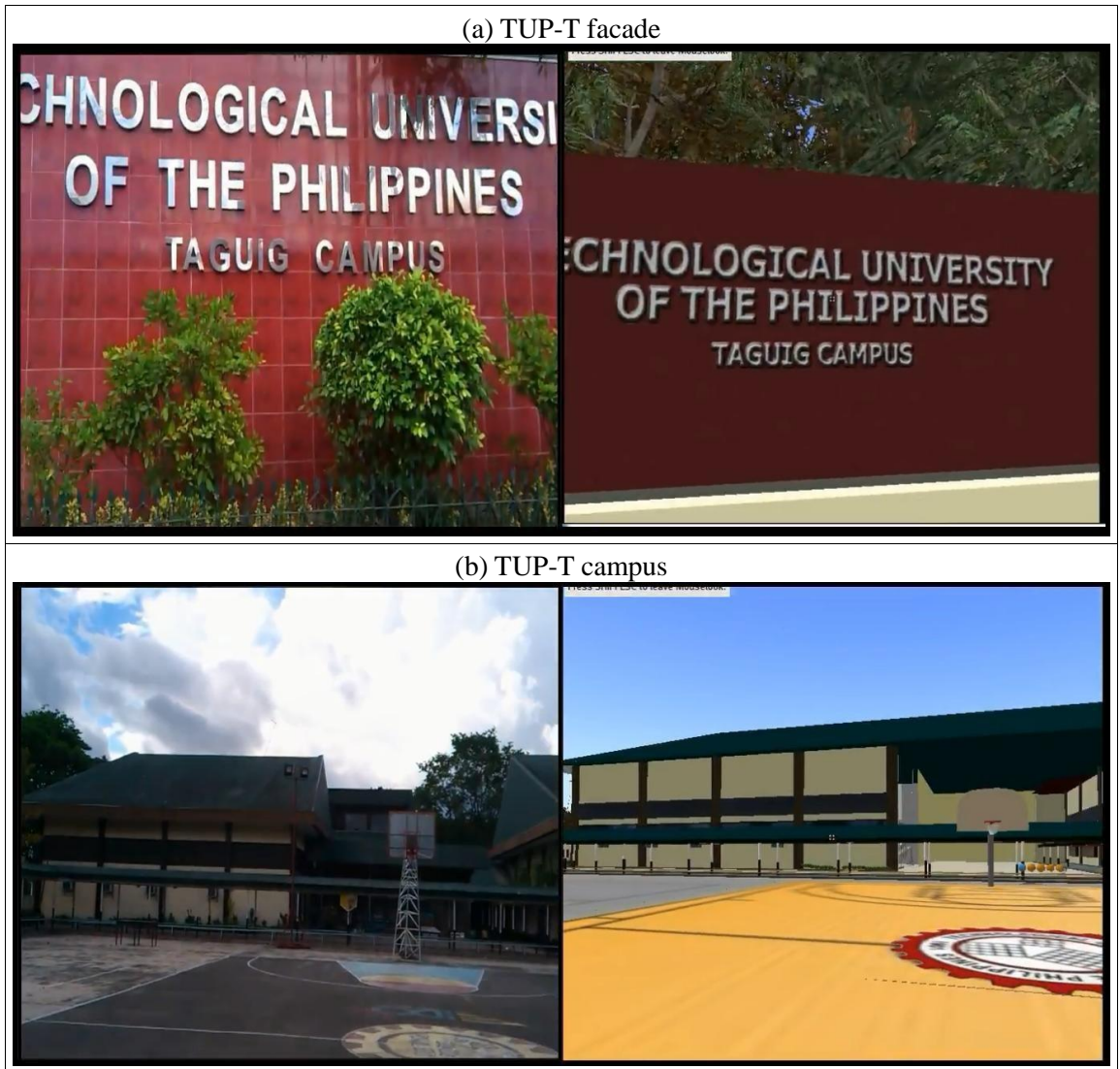

**Figure 5**. The TUP-T (a) facade, and (b) campus as seen in photographs (left) and as converted into virtual objects and seen through CVLV (right).

*E. Audio Chat Tool*

We used the voice chat software suite called Mumble & Murmur to provide our students the capability to interact and orally communicate with each other and with their instructor. Mumble is the client software system and is installed in the students' computers, while Murmur is the server software system and is installed in the laboratory's server computer. The software suite is FOSS and performs audio communication with low-latency yet high quality voice chat capability through the Opus interactive audio codec [22]. Opus is the standard for speech and music transmission over the Internet as specified in the RFC 6716 by the Internet Engineering Task Force (IETF). Opus is a combination of the Speex and the Constrained Energy Lapped Transform (CELT) ultra-low delay audio codecs. Both Speex and CELT are audio compression and transmission algorithms. The only difference is that Speex was optimized for speech [23], while CELT was optimized for music [24]. Since Opus is a combination of the two codes, then Opus is optimized for both speech and music transmission over the Internet and therefore is a very good codec for interactive voice communication.

Aside from the standard audio transmission characteristics of Mumble, we chose the suite because of the additional pre-processing it performs to remove background noise. The removal of the background noise improves the clarity of spoken sound. Additionally for stereo sound systems which we implemented for realistic sound experience of the students, Mumble also performs positional audio [25] which allows the realistic sound rendering of the position of the source of sound. For example, if Student A is talking to the avatar of Student B who is standing at Student A's right side, the audio of Student B will be louder at Student A's right speaker. If Student A leaves the virtual classroom, her audio will not be heard by the students whose respective avatars were left in the virtual classroom. Without positional audio capability, all students will hear each other's spoken words with the same loudness and at the same perceived direction, even if the avatar representing the student is far from the other students.

## HARDWARE PLATFORM

*A. Client Desktop Computers*

In our 3D-AVLE hardware subsystem, we used 30 desktop computers with identical motherboards, processors, memory capacities and operating systems. We wanted our students to experience the same satisfaction, operational speed, and perceived contentment with regards to learning and teaching sessions conducted in our 3D-AVLE. Specifically, our client desktops are equipped with Intel Core 2 Duo processor, 2GB DDR3 memory, and Intel G41 Express chipset with built-in video card. These desktops are the ones available in the TUP-T computer laboratory. Our hardware system can handle 30 to 32 simultaneous online users. We measured the response rate of our 3D-AVLE, particularly the smoothness of the rendering of the virtual world viewer, and we found that it can handle the load demand at maximum capacity. Even though our hardware system can handle up to 32 concurrent users, we chose to connect up to 16 students only to provide us with an allowance for a possible backlog that we did not foresee in this optimization step.

*B. Communication Platform*

We provided the students each with a headset and a microphone which they used as their oral communication tool. This hardware allowed the students to participate in class discussion, raise queries to instructors or fellow students, answer questions themselves, participate in a debate, or voice out their own opinion in the subject being delivered. With a press of a button, CVLV allows a student to "shout," "talk," or "whisper" [7, 20]. When the "shout" button is pressed, a word spoken with a normal loudness by a student will be heard with a louder volume in the 3D-AVLE by all students whose respective avatars are in the same room, the loudness of which is corrected by the distance of the user's avatar to the other avatars. A "talk" means that the spoken words of the user will only be heard by those users whose avatars are in close proximity to hers. A "whisper" allows a user to directly talk to a selected avatar, and only the person representing that avatar will be able to hear the spoken words by the user, even if in reality the user actually shouted those words.

## IMPLEMENTATION EXPERIENCES

*A. Access to 3D-AVLE*

Like any other multi-user applications, each student needs to undergo registration first before they can be allowed to participate in the 3D-AVLE. The students needed to sign-in with their Firstname and Lastname on the respective text boxes provided by 3D-AVLE in the registration box. For security purposes, students set their own respective passwords. After the registration, they started with the creation of their own avatar.

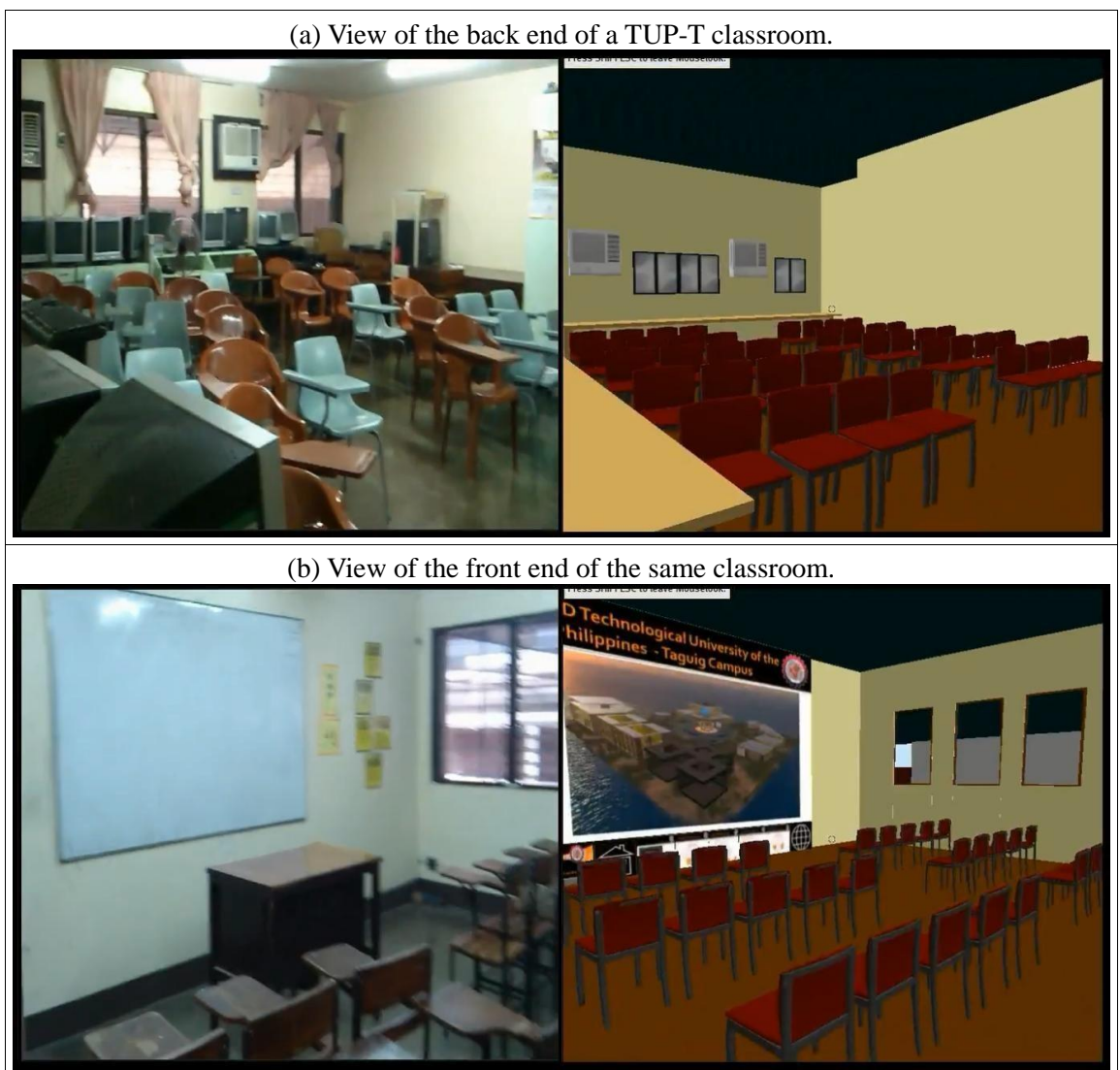

**Figure 6**. The back and front end views of a TUP-T classroom as seen in photographs (left) and as converted into virtual objects and seen through CVLV (right).

We allowed our students to change their avatar appearances as they please by providing then with a wide variety of choices for clothes and accessories. As soon as their accounts were created, we then proceeded in setting up the things needed to start the lesson in the virtual classroom. Figure 6 shows the back and front views of a real TUP-T classroom and its virtual counterpart in the 3D-AVLE.

*B. Preparation of Teaching Materials*

We needed to upload the images of the presentation slides that we used to deliver our subject matter. The slide images were transferred from the OpenSim inventory to Sloodle [26] using a drag-and-drop process. Sloodle is a presentation slides presenter used primarily for OpenSim. With the presenter, the presentation slides was translated into an object with a script to display the slides using navigational buttons.

We experienced permission lock-out while using Sloodle, which maintains a list of students that we allowed to view our presentation slides. Initially, we provided permission to 10 students without problem. However, when we added the number of students to be allowed access to the slides, the problem surfaced when the presenter need to accept the request to view the presentation from the 13th to 15th random users. Here, CVLV hanged up after the permission window showed up. This affected the rest of the students who were already given permission. To resolve this, students who were included in the permitted list needed to log-out of the 3D-AVLE leaving behind those that need to be set-up. As soon as the request for the remaining students was accepted, the problem was solved.

*C. Measurement of Student Perception*

We modified the Athabasca University LimeSurvey Community of Inquiry (COI) framework for Virtual Worlds to provide us a measure for assessing the students' perception in using the 3D-AVLE. COI is a survey form developed to allow researchers, teachers and students to assess the learning that occurs in SL/TSL [27, 28]. We customized COI for use by the TUP-T students. The COI contains three categories: Teaching Presence (TP), Social Presence (SP), and Cognitive Presence (CP). In order to measure numerically the different elements of the COI from a student's perspective, we used the five-point Likert scale instrument [29-31] as follows: 1 means Strongly Disagree (SD), 2 means Disagree (D), 3 means Neutral (N), 4 means Agree (A), and 5 means Strongly Agree (SA). These numbers estimate the perception of the virtual learners [32-33] as they answer their respective levels of agreements to the respective statements in the COI. We conducted the survey after the students took their second examination and the results show that the 3D-AVLE can provide the students with the teaching, social, and cognitive presence that they need for learning. We presented this result in detail in another paper [8].

*D. Measurement of Student Performance*

To measure the performance of a student in a 3D virtual environment, we conducted pretest and posttest examinations during their 3D-AVLE learning experiences. The measurement is based on the framework shown in Figure 7. Here, we divided our classes into two groups, Group A and Group B. Each group is composed of up to 15 students. For each lesson, we conducted a pretest which all students from both groups took. The pretest questions asked were the same for both groups. The students in Group A proceeded to take the lesson conducted in 3D-AVLE while those in Group B took the same lecture in a traditional classroom setting. Students from both groups, then, took the post-test examination. We repeated this process with 3 more other lessons. However, in succeeding lessons, students from the two groups interchanged. The students who took the immediate past preceding lesson from 3D-AVLE took the immediate next succeeding lesson from the traditional classroom. Similarly, those students who took the immediate past preceding lesson from the traditional setting took the immediate next succeeding lesson from 3D-AVLE. Results show that students who took their lessons from 3D-AVLE performed better in the post-test while all pretest scores for both groups are not statistically different from each other [9]. This result is in agreement with what was found earlier by other researchers [13].

**SUMMARY**

In this paper, we presented our experiences in creating and implementing a 3D-AVLE for TUP-T. We discussed the particular software that we used and the corresponding reasons why we chose them. We presented our hardware platform and how we optimized them for use by up to 16 students who concurrently connected to be able to attend a virtual lesson. We also presented our implementation experiences, including those activities where we measured the student perception on using the 3D-AVLE, as well as their performance in taking lessons with it. Students perceived that the 3D-AVLE was successful in providing them with teaching, social, and cognitive presence [8], and at the same time, using it for learning increases student performance [9].

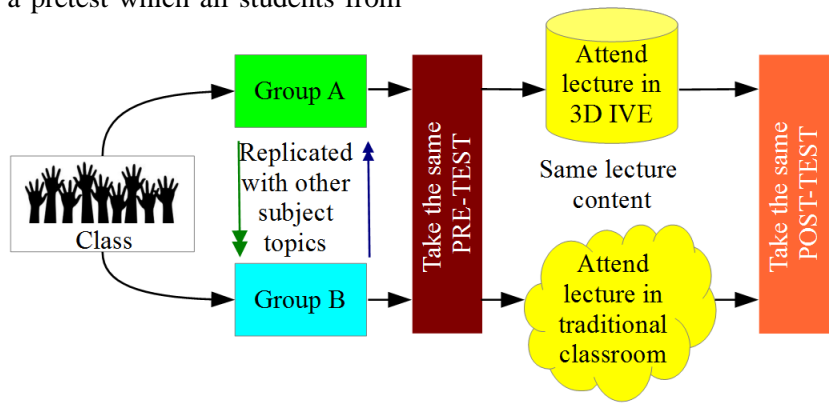

**Figure 7**. The conceptual framework for measuring the performance of students in 3D-AVLE.

**ACKNOWLEDGEMENTS, ETHICAL CONSIDERATIONS, AND AUTHOR CONTRIBUTIONS**

This research endeavor was funded by the Department of Science and Technology (DOST) Advanced Science and Technology Human Resource Development Program (ASTHRDP) Graduate Scholarship with M.V. Rivera and N.R. Valdez as scholar-grantees under the Master of Information Technology (MIT) program of the Graduate School (GS) of the University of the Philippines Los Baños (UPLB) with J.P. Pabico as their senior research adviser. The hardware subsystems used for the 3D-AVLE are from the TUP-T, as well as the subject students.

The Ethics Committee of TUP-T approved the conduct of this research on the students as human subjects of research. The authors declare that the academic and social well beings of the students were respective during the conduct of the research.

The following are the respective contributions of the authors: N.R. Valdez assembled and optimized the hardware subsystem of the 3D-AVLE, conducted the real-world observation and statistical analyses of the student performance experiments, and prepared the final paper; M.V. Rivera assembled and optimized the software subsystem of the 3D-AVLE, implemented the social experiments and statistical analyses on student perception, and prepared the final paper; and J.P. Pabico formulated the hardware and software problems, provided the requirements for the solution to the problems, interpreted the statistical analyses, and edited the final manuscript.

All authors declare no conflict of interest.

*Note: The use of the female gender in this text is just a writing style without prejudice to the other.